\documentclass{qjmam}
\usepackage{amssymb}
\usepackage{amsmath}
\graphicspath{{./figures/}}
\usepackage{cleveref}
\usepackage{color}

\startpage{1}
\yr{2019}
\vol{52}
\issue{1}

\newcommand{\pfr}[2]{\ensuremath{\frac{\partial #1}{\partial #2}}}
 
\newcommand{\dd}{\mathrm{d}}

\newcommand{\Rey}{Re}

\newcommand{\beq}{\begin{equation}}
\newcommand{\eeq}{\end{equation}}

\DeclareMathAlphabet\mathbit
    \encodingdefault\rmdefault\bfdefault\itdefault
\DeclareOldFontCommand{\bi}{\normalfont\bfseries\itshape}{\mathbit}

\newcommand{\be}{\begin{equation}}
\newcommand{\ee}{\end{equation}}

\def\fakebold#1{\relax\ifvmode\leavevmode\fi%
\ifmmode%
\setbox0=\hbox{$#1$}%
\else%
\setbox0=\hbox{#1}%
\fi%
\kern-.02em\copy0 \kern-\wd0%
\kern .04em\copy0 \kern-\wd0%
\kern-.0125em\raise.02em\box0%
}%

\renewcommand{\geq}{\geqslant}
\renewcommand{\leq}{\leqslant}

\begin{document}

\title[{Flow induced by half-line sources}] {A note on viscous flow induced by half-line sources bounded by conical surfaces}

\author[P.~Rajamanickam \& A.~D.~Weiss] {Prabakaran Rajamanickam \and Adam D. Weiss}

\address{Dept. of Mechanical and Aerospace Engineering, University of California San Diego,\\
La Jolla, CA {\rm 92093}, USA}

\received{\recd 25 July 2019. \revd }

\maketitle

\eqnobysec

\begin{abstract} 
 In this paper axisymmetric solutions of the Navier-Stokes equations governing the flow induced by a half-line source when the fluid domain is bounded by a conical wall are discussed. Two types of boundary conditions are identified; one in which the radial velocity along the axis is prescribed, and the other in which the radial velocity along the axis is obtained as an eigenvalue of the problem. The existence of these solutions are limited to a range of Reynolds numbers and the transition from one case to the other are discussed in detail.
\end{abstract}

\section{Introduction}

 Conically similar flows have attracted many researchers in the past and the present ever since the discovery of a new Navier-Stokes solution by Landau~\cite{landau1944} and Squire~\cite{squire1951}, expressed in spherical-polar coordinates. The so-called Landau-Squire jet describes the motion of an unbounded fluid driven by a momentum flux of singular nature enforced at the origin. Unfortunately, solutions when the fluid domain is bounded by a solid surface which has conical symmetry have not been found. Nevertheless, there are exhaustive collections of viscous solutions that do not satisfy the no-slip condition at the wall. The no-slip condition was found to be satisfied if there is a transpiration across the wall~\cite{morgan1956} or if one permits the polar axis to be singular.  The extensive review can be found in the book by Drazin and Riley~\cite[pp. 78-88]{drazin2006} and the references therein. Besides solutions of the Navier-stokes, conical similarity is a feature utilized for obtaining a wide family of solutions to the heat, diffusion, and magneto-hydrodynamic equations as discussed, for example, in the work of Shtern~\cite{shtern2012}.

The exact solution with a singular axis was discovered by Goldshtik~\cite{goldshtik1960} and Serrin~\cite{serrin1972}, where a half-line vortex at the axis was chosen to serve the purpose. However, the corresponding problem with the line vortex replaced by a line source or sink drew very little attention. The flow induced by half-line volume sources/sinks bounded by a plane wall was studied by three relatively unknown works, one by Golubinskii and Sychev~\cite{golubinskii1976} and the others by Goldshtik and Shtern~\cite{goldshtik1988,goldshtik1990}. The fluid motion induced by line-sources and line-sinks was found to be essentially different from each other and  existence of such solutions also found not to be guaranteed for all Reynolds number. The work by Schneider~\cite{schneider1981} should also be mentioned here, where only a half-line sink was considered as a model for flow entrained by axisymmetric jets and plumes. Half-line sources or sinks in an unbounded fluid domain are discussed in~\cite{pillow1985,paull1985}.

The purpose of this paper is to generalize the works of~\cite{golubinskii1976,goldshtik1988} to solid walls that are not just planes perpendicular to the polar axis and to provide additional clarifications to the problem. In order to do that, two different cases are considered, case I and case II, that differ from each other by one boundary condition. Both cases, to be defined in the next section, pertain to flow induced by half-line sources located along the positive side of the polar axis. Depending on the problem parameters, we identify in general two types of fluid motion, namely \textit{pure outflow} and \textit{reversed flow}. By pure outflow, we mean that the velocity component in the radial direction is nonnegative everywhere. In the reversed flow configuration, this velocity component undergoes a sign change within the fluid domain. As will be seen, the solution also exhibits non-uniqueness for a given Reynolds number, which is a typical characteristic of conically similar solutions. This non-uniqueness and the flow separation are already apparent in flows induced by the line vortices~\cite{shtern1996,shtern1999}.

\section{The governing equations and the boundary conditions}

Consider a half-line source situated along the positive part of the polar axis, that ejects fluid radially outwards with a volumetric flow rate per unit axial length $Q$ that is constant. The fluid domain is bounded by a solid conical wall whose axis is supposed to be along the polar axis with cone vertex located at the origin. Let $\alpha$ be the angle that the solid wall makes with the line source.

The convenient choice of coordinates are the spherical coordinates $(r,\theta,\phi)$ that defines the velocity field $(v_r,v_\theta,0)$ which are assumed to be functions of $r$ and $\theta$ only. The motion on the axial plane is described by a Stokes stream function $\psi$ such that 
\begin{equation}
    v_r =\frac{1}{r^2\sin\theta} \pfr{\psi}{\theta} \quad \text{and} \quad  v_\theta= -\frac{1}{r\sin\theta} \pfr{\psi}{r}.
\end{equation}
Introducing $\xi=\frac{1}{2}(1-\cos\theta)$ in place of $\theta$ and seeking solutions of the form $\psi =  Q r f(\xi)$, the axisymmetric Navier-Stokes equation without swirl can be shown to reduce to
\begin{equation}
    \xi(1-\xi)f^{iv} + 2(1-2\xi) f''' + 2\Rey(ff''' + 3 f' f'') = 0, \label{fourthorder}
\end{equation}
where we have defined the Reynolds number to be $\Rey=Q/(4\nu)$. Integrating the foregoing equation twice, we obtain
\begin{equation} \label{secondorder}
    \xi(1-\xi) f'' + 2 f + 2\Rey f f' = 2a_1\xi + a_2,
\end{equation}
which upon further integration places the equation in Riccati form, first derived by Slezkin~\cite{slezkin1934},
\begin{equation}
    \xi(1-\xi) f' - (1-2\xi) f + \Rey f^2 = a_1 \xi^2 + a_2 \xi + a_3, \label{Riccati}
\end{equation}
where $a_1$, $a_2$ and $a_3$ are integration constants. 

This work considers two sets of boundary conditions. While the boundary conditions at the conical wall
\begin{equation}
    f(\xi_w)=f'(\xi_w)=0,\label{bcw}
\end{equation}
are the same for both cases, the boundary conditions at the axis are given by
\begin{align}
    \text{case I}: \quad &f+1 = \sqrt{\xi}f'' = 0\quad &&\text{as}\quad \xi\rightarrow 0, \label{bcaI}\\  
    \text{case II}: \quad &f+1 = f' - M= 0\quad &&\text{as} \quad  \xi\rightarrow 0, \label{bcaII}
\end{align}
where $|M|< \infty$ is a known number. In case I, a symmetry condition on the axial velocity is imposed, whereas in case II, $v_r$ is assumed to be given. The boundary condition on the polar velocity is the same in both cases. These boundary conditions evidently determine the integration constants defined earlier for both cases,
\begin{align}
     a_3 = 1 + \Rey, \quad a_2 = -2[1+\Rey f'(0)], \quad a_1 = - (a_3 + a_2 \xi_w)/\xi_w^2. \label{constants}
\end{align}

For $f'(0)$ to be bounded, it is required that the $\lim_{\xi\rightarrow 0}\xi f''(\xi)$ vanishes identically, for otherwise $f'(0)$ will become unbounded in logarithmic fashion. The symmetry condition imposed at the axis in case I is just a sufficiency condition for $v_r$ to be bounded, but not necessary. In addition, the relation between the wall stress $f''(\xi_w)$ and the axial velocity $f'(0)$ is given by
\begin{equation}
    f''(\xi_w) = - \frac{2}{\xi_w^2} - \frac{2\Rey [1-f'(0)\xi_w]}{\xi_w^2(1-\xi_w)}, \label{fwwfa}
\end{equation}
obtained directly from \eqref{secondorder} evaluated at $\xi=\xi_w$.  For case I, the solution provides in particular the value of $f'(0)$ thereby determining the wall shear stress through \eqref{fwwfa} whereas in case II, the wall stress is known a priori.

\begin{figure}
\centering
\includegraphics[scale=0.55]{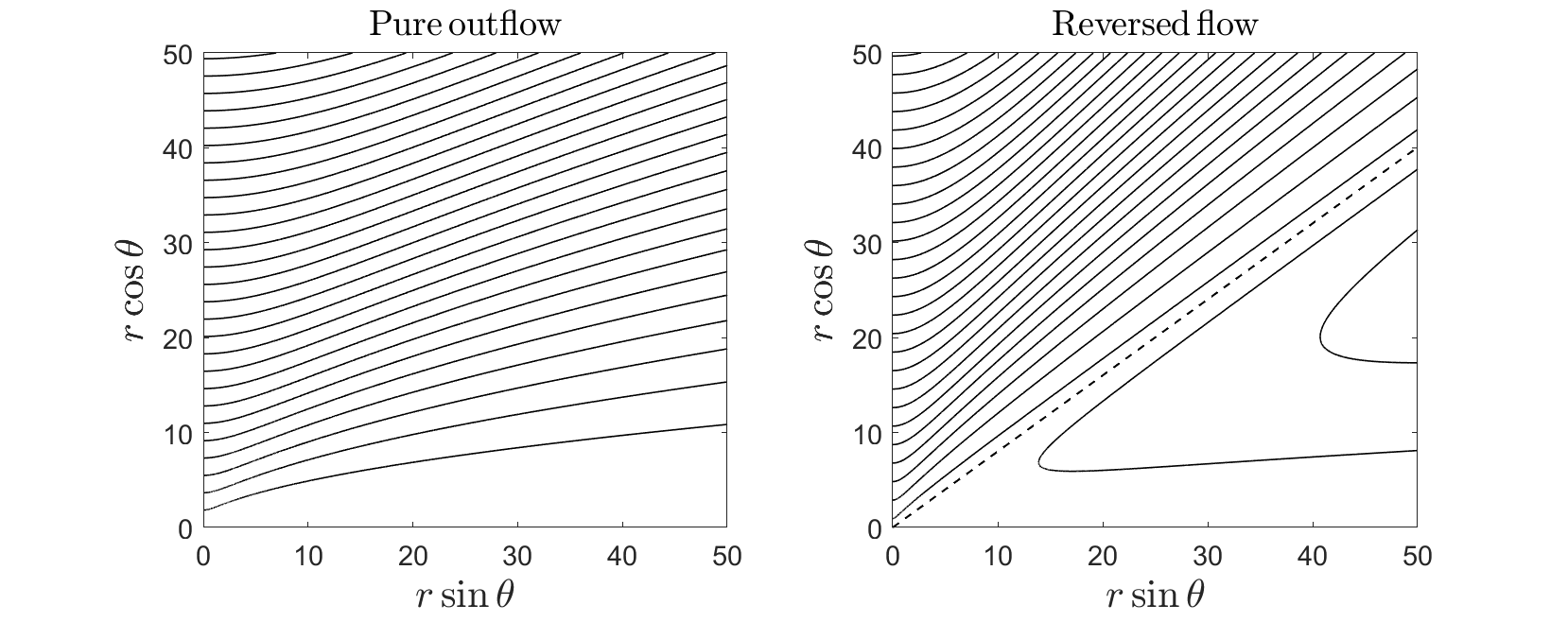}
\caption{Equi-spaced streamlines plotted for a plane wall, $\alpha=\pi/2$ with two sample results computed for case I with (left) $\Rey = 0.1$ and (right) $\Rey = 0.215$, one corresponding to pure outflow and the other to reversed flow.}
\label{fig:contour}
\end{figure}

The numerical solution of the problem may proceed in a variety of ways. For instance, the first order equation \eqref{Riccati} may be used, as has been done in \cite{goldshtik1988,schneider1981}, where in addition for case I an iterative method is required to determine the unknown eigenvalue $f'(0)$. Alternatively, as has been done in the present analysis, the fourth order equation \eqref{fourthorder} can be solved directly by means of Newton's iteration. As will be seen below there is a duplicity of solutions for a given $\Rey$ in case I, which could be computed by treating $\Rey$ as an eigenvalue for a given value of $f'(0)$.

\section{General characteristics of solution}
\label{sec:general}

Irrespective of the two cases defined in the previous section, there exists only two types of flow fields for all possible parameters. As defined in the introduction, in pure outflow, $f(\xi)<0$ everywhere in the domain, whereas in the reversed flow $f(\xi)$ changes sign within the fluid domain. The flow reversal occurs near the solid wall where fluid from infinity is directed inwards and reversed backwards after reaching the origin, whereas the fluid ejected from the axis is directed upwards. The sample streamlines plotted in Fig.~\ref{fig:contour} for $\alpha = \pi/2$ with $\Rey = 0.1$ and $\Rey = 0.215$, respectively, give an idea of the typical fluid motion. The dashed line in the figure corresponding to the point $f(\xi)=0$ represents the cone that separates the slow-recirculation region from the fluid ejected rapidly from the axis. Since reversed flow always occurs near the wall, the fluid motion can also be distinguished between pure outflow and the existence of reversed flow by observing the second derivative $f''(\xi_w)$, with $f''(\xi_w) >0$ corresponding to reverse flow and $f''(\xi_w)<0$ with pure outflow. The angle of the separating cone decreases as the Reynolds number increases. In the limiting case, where the separating cone degenerates to the axis, we have a jet at the axis with infinite velocity gradient and the recirculating region occupies the whole domain. In the reversed flow solution, the function $f$ is found to be very sensitive to the changes in the values of $\mathit{Re}$ and takes large positive values as $\Rey$ increases, making it harder to integrate the equations numerically.


 \begin{figure}
\centering
\includegraphics[scale=0.6]{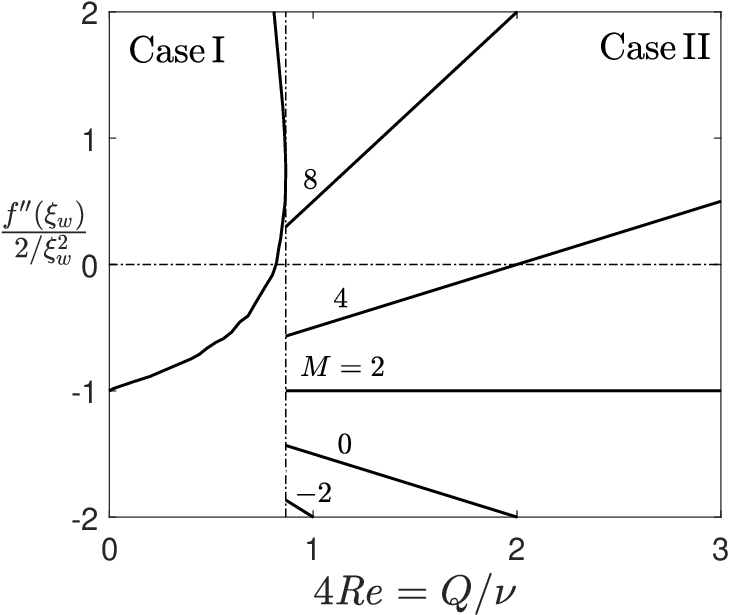}
\caption{The scaled wall stress as a function of Reynolds number for case I and case II for the plane wall problem, $\alpha=\pi/2$.}
\label{fig:general}
\end{figure}

Replacing in case I the link source by a sink (i.e $f=1$ at $\xi = 0$) one obtains the flow studied by~\cite{schneider1981} for which solutions existed for all Reynolds numbers. On the other hand, for a line source we shall show below that solutions only exist for Reynolds number below a critical value $\Rey=\Rey_c$. For $\Rey<\Rey_c$, $f'(0)$ is obtained as an eigenvalue of the problem thereby determining the wall shear stress $f''(\xi_w)$ through \eqref{fwwfa}. The wall stress is plotted as function of Reynolds number in Fig.~\ref{fig:general} for $\alpha = \pi/2$ exhibiting a duplicity of solutions in the range $0<\Rey<\Rey_c$. As said before, both branches can be obtained by treating $\Rey$ as an eigenvalue for a given value of $f''(\xi_w)$. The upper branch always corresponds to reversed flow and is an unstable branch as can be seen from looking at the trend for small Reynolds numbers thereby requiring no further discussion. 

\begin{figure}
\centering
\includegraphics[scale=0.6]{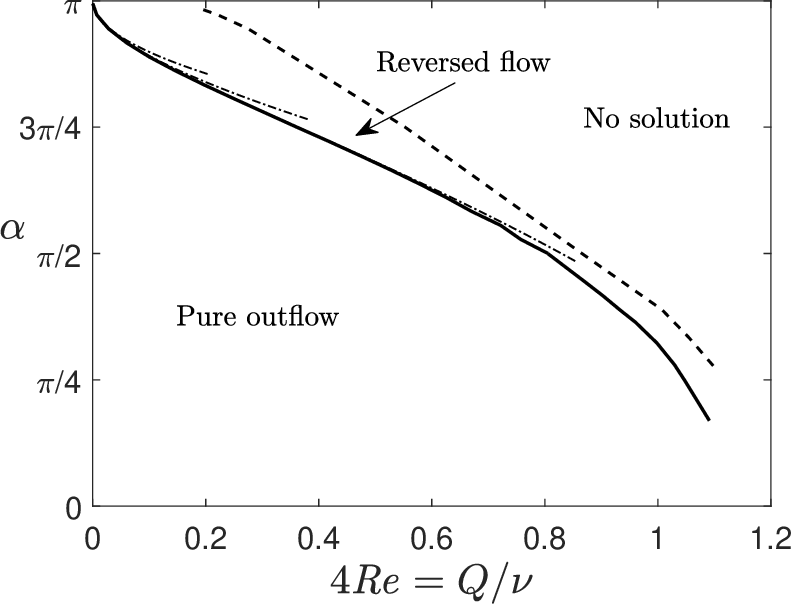}
\caption{Critical Reynolds numbers $Re_s$ (solid line) and $Re_c$ (dashed line) as functions of cone-angle $\alpha$. The dash-dotted lines correspond to the predictions presented in \S~\ref{sec:critical}.}
\label{fig:critical}
\end{figure}

Before reaching the point $\Rey=\Rey_c$, there exists another critical value $\Rey=\Rey_s$, where $f(\xi)$ undergoes a sign change within the fluid domain and where $f''(\xi_w)=0$. The critical value $\Rey_s$ at which $f''(\xi_w)$ vanishes can be obtained numerically by treating $\Rey$ as an eigenvalue and imposing the condition $f''(\xi_w)=0$. These critical values are plotted as a function of the cone angle $\alpha$ in Fig.~\ref{fig:critical} as solid lines.  As $\alpha\rightarrow \pi$, the critical Reynolds number becomes zero because the geometry is such that it allows for flow separation from the wall easily. Decreasing $\alpha$ evidently postpones the bifurcation to larger Reynolds number and the trend is monotonic. Estimation of these critical values is given in \S~\ref{sec:critical}. In the same figure, the second critical Reynolds number $\Rey_c$ separating case I and case II is displayed as a dashed line. The reversed flow region in case I can occur as can be seen for narrow range of Reynolds number only, i.e., in the range $\Rey_s\leq \Rey\leq \Rey_c$.

For $\Rey>\Rey_c$, the fact that we are unable to find any $f'(0)$ for a given $\Rey$ satisfying the boundary conditions for case I means either there exists no value of $f'(0)$ for that range or there exist solutions for any value of $f'(0)$. The latter observation is in fact true and solution can be obtained for any values of $f'(0)$ that belongs to case II. 

Let us now consider case II. The curves of $f''(\xi_w)$ vs. $\Rey$ are just straight lines with slopes determined by $M$ given by \eqref{fwwfa}. It follows that $f''(\xi_w)$ does not depend on the Reynolds number and takes the value $-2/\xi_w^2$ when $M=1/\xi_w$ (see the $M=2$ line for the case $\xi_w=1/2$ in Fig.~\ref{fig:general}). This condition then defines different possible motions for the fluid. For $M\leq 1/\xi_w$ including the case $M=0$, flow reversal never occurs for any $\Rey$ and as $\Rey\rightarrow \infty$, the flow splits into an outer potential flow and a near-wall boundary layer. These structures are discussed in \S~\ref{sec:boundarylayer}. When $M>1/\xi_w$, the flow may begin as pure outflow for $\Rey>\Rey_c$, but eventually always becomes a reversed flow at $\Rey=(1-\xi_w)/[2(M\xi_w-1)]$. However, the flow is always reversed for $M\geq 1/\xi_w + (1-\xi_w)/\Rey_c$.

\section{Explicit solutions and limiting behaviours}
\label{sec:explicit}

\subsection{Yatseyev solution}
The general solution of the Riccati equation~\eqref{Riccati} is due to Yatseyev~\cite{yatseyev1950}, who used the transformation $f=\Rey^{-1} \xi(1-\xi)u'/u$ to convert the Riccati equation to a linear second-order differential equation,
\begin{equation}
    u'' - \frac{\Rey(a_1\xi^2 + a_2\xi + a_3)}{\xi^2(1-\xi)^2} u = 0.
\end{equation}
The point $\xi=0$ is a regular singular point of the equation. The local behaviour of the solution near the origin is of the form $u\sim \xi^{c/2}$, where $c=-2\Rey,\,2(1+\Rey)$. To satisfy the condition $f(0)=-1$, it is required that $c=-2\Rey$, although keeping the solution corresponding to $c=2(1+\Rey)$ is harmless. 

The equation can be converted to the hypergeometric differential equation by introducing the substitution $u=\xi^{c/2}(1-\xi)^{(a+b-c+1)/2}w(\xi)$, where
\begin{align}
    a &= \frac{c}{2} \pm \frac{1}{2}\left(\sqrt{1+4\Rey(a_1+a_2+a_3)}+\sqrt{1+4\Rey a_1}\right),\\
    b &= \frac{c}{2} \pm \frac{1}{2}\left(\sqrt{1+4\Rey(a_1+a_2+a_3)}-\sqrt{1+4\Rey a_1}\right),\\
    c&=-2\Rey,\, 2(1+\Rey).
\end{align}
The solution behaviour near the axis is independent of the choice made for $c$ and we shall choose $c=-2\Rey$. Then the full solution with a regular singular point $\xi=0$ can be written as
\begin{equation}
    u(\xi) = \xi^{c/2}(1-\xi)^{(a+b-c+1)/2}[ {}_2 F_1(a,b;c;\xi)+k \xi^{1-c}{}_2 F_1(a-c+1,b-c+1;2-c;\xi)], \label{hyper}
\end{equation}
where ${}_2 F_1$ is Gauss' hypergeometric function and the constant of integration $k$ can be obtained by imposing the wall boundary condition. The above solution fails when the Reynolds number is a half-integer or an integer, i.e., when $\Rey=n/2$, where $n=0,1,2,3,..$. In these special cases explicit solution can be written~\cite{DLMF}, although we do not present them here. We develop explicit solutions only to the extent that reveals the limiting form for small $\xi$; otherwise the governing equations are integrated numerically.

For any Reynolds number other than that those for which~\eqref{hyper} becomes invalid as $\xi\rightarrow 0$, it can be verified that $f=-1+f'(0)\xi+\cdots$. In evaluating the next term in the series, care is required. In the range $0<\Rey<1/2$ (the range where transition between case I and II lies), we notice that the second term in~\eqref{hyper} do not contribute to the expansion for small $\xi$ till the quadratic term and therefore, we can write for that range,
\begin{equation}
    f=-1+f'(0)\xi + \frac{\Lambda}{1-2\Rey}\xi^2 + \cdots,
\end{equation}
where $\Lambda = a_1 - f'(0)[1+\Rey f'(0)]$ or
\begin{equation}
    \Lambda = \frac{1}{\xi_w^2}[-f'(0)\xi_w^2-\Rey[f'(0)]^2\xi_w^2 + 2\Rey f'(0)\xi_w + 2\xi_w -1 -\Rey]. \label{Lambda}
\end{equation}
The choice of sign for $a$ and $b$ is immaterial in obtaining the expansion for $f(\xi)$. This implies that whenever $\Lambda\neq 0$, the symmetry condition $\lim_{\xi\rightarrow 0}\sqrt{\xi} f''(\xi)$ is always satisfied for $\Rey<1/2$.  The fact that the symmetry condition is satisfied in both cases, but $f'(0)$ is determined as an eigenvalue in case I and $f'(0)$ is assumed to take any value in case II contradicts each other and it follows that case I and case II cannot exist together in the range $0<\Rey<1/2$, provided $\Lambda\neq 0$. What happens when $\Lambda=0$ is discussed below.

\subsection{Explicit solution obtained from particular solution for $\Lambda=0$}

 \begin{figure}
\centering
\includegraphics[scale=0.6]{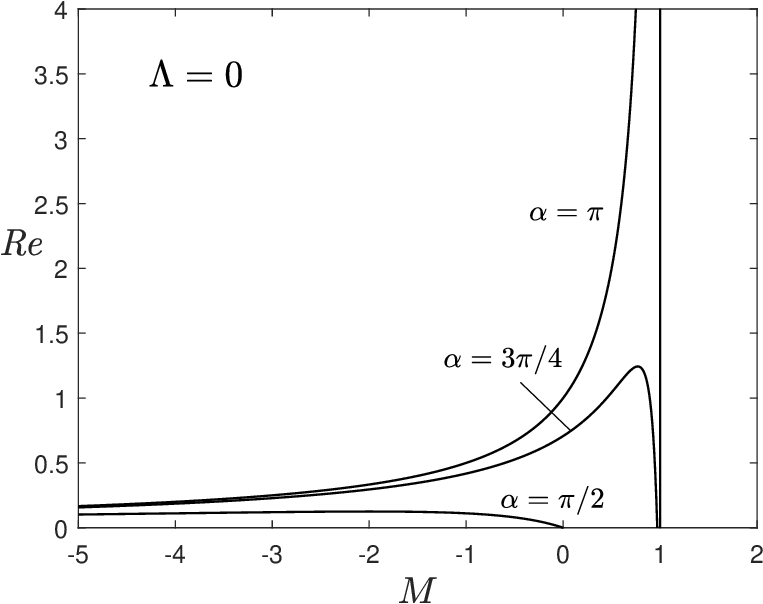}
\caption{$\Lambda = 0$ contours in the $\Rey$ - $M$ plane for several $\alpha$ as given by \eqref{Lambda}. }
\label{fig:lambda}
\end{figure}

The general solution of the Riccati equation can be constructed if a particular solution is known. It is a matter of fact that the Riccati equation~\eqref{Riccati} admits the particular solution $f(\xi) = -1 + M\xi$ whenever $\Lambda=0$. In this section, we shall exclusively use $M$ instead of $f'(0)$ simply because $\Lambda=0$ and case I are not compatible with each other as we shall see. The general solution can be built by anticipating the solution $f(\xi) = -1 + M\xi + 1/v(\xi)$, with $v(\xi)$ satisfying the equation
\begin{equation}
    \xi(1-\xi) v' + \{1 + 2\Rey- 2x [1+\Rey M]\} v = \Rey.
\end{equation}
The solution that satisfies the boundary conditions is
\begin{equation} 
   v(\xi) = \frac{1}{p}\left(\frac{1}{1-\xi}\right)^{1-2\Rey(1-M)} \left(\frac{1}{\xi}\right)^{1+2\Rey} + \frac{\Rey (1-\xi)^{-1+2\Rey(1-M)}}{\xi^{1+2\Rey}}\int_{\xi_w}^\xi \frac{x^{2\Rey} \, \dd x }{ (1-x)^{-2\Rey(1-M)}},
\end{equation}
where $p=[1-M\xi_w]\xi_w^{-1-2\Rey}(1-\xi_w)^{-1+2\Rey(1-M)}$ have been introduced for brevity. Moreover, the integral originated from the inhomogeneous term can be written in terms of the incomplete beta function. From the above solution, we conclude that for small $\xi$,
\begin{equation}
    f=-1+M\xi + p \xi^{1+2\Rey} + \frac{p^2\Rey}{1+2\Rey}\xi^{2+2\Rey} + \cdots
\end{equation}
Note that the constant $p$ cannot be zero unless $\xi_w=1$, a singular case that we avoid here, if the condition $\Lambda=0$ is to be met. This leads to the conclusion that while the boundedness condition $\lim_{\xi\rightarrow 0}\xi f''(\xi)=0$ is satisfied for all Reynolds number, the satisfaction of symmetry condition $\lim_{\xi\rightarrow 0}\sqrt{\xi}f''(\xi)=0$ is not possible for $\Rey<1/4$.

It is also interesting to understand in the parametric space what $\Lambda=0$ means. Fixing different values for $\xi_w$, the curves traced by the condition $\Lambda=0$ are plotted in the $\Rey$-$M$ plane in Fig.~\ref{fig:lambda}. The equation $\Lambda=0$ does not yield any positive roots for the Reynolds number for values of cone angles less than $\alpha\lesssim\pi/4$. It is also clear from Fig.~\ref{fig:critical} that for most $\alpha$ (typically larger than $\pi/4$) the critical Reynolds number $\Rey_c$ is generally less than $1/4$. Since the symmetry condition is not satisfied for $\Rey<1/4$ and for $\Rey>1/4$ (small cone-angles in case I), $\Lambda=0$ do not have any solution, we can conclude that $\Lambda=0$ and case I are exclusive. Moreover, the curve in $\Rey$ vs. $M$ plane calculated from case I for $\alpha=\pi/2$ is still larger than the curve given in Fig.~\ref{fig:lambda}. Thus, in considering any-angles larger than $\pi/2$, again we can conclude that $\Lambda=0$ will necessarily correspond to case II, but not case I.

Since it is shown that $\Lambda=0$ does not correspond to case I and since it has been proven that the symmetry condition is always satisfied when $\Rey<1/2$, the conclusion to be drawn is that case II does not exist in the range $0\leq \Rey \leq \Rey_c$. One can easily verify that there is no solution to the governing equation when $\Rey=0$ that satisfies all boundary conditions required by case II.

\subsection{Limiting behaviour of the solution near the axis}

The results for the limiting forms obtained from the explicit solutions can also be derived to an extent by introducing $f=-1+f'(0)\xi + g(\xi)$ with $|g|\ll 1$ to obtain a local solution near the axis. The function $g(\xi)$ as $\xi\rightarrow 0$ can be shown to satisfy the following differential equation
\begin{equation}
    \xi g' - (1+2\Rey) g = \Lambda \xi^2.
\end{equation}
The general solution is
\begin{align}
    \Rey \neq \frac{1}{2}: &\quad g = c_1 \xi^{1+2\Rey} + \frac{\Lambda \xi^2}{1-2\Rey},\\
    \Rey = \frac{1}{2}: &\quad g = c_1 \xi^2 + \Lambda \xi^2\ln \xi.
\end{align}
By comparing with the limiting forms obtained earlier, we can argue that $c_1=0$ for $\Rey<1/2$ and equal to $p$ when $\Lambda=0$. For $\Rey>1/2$, since $g\sim \xi^2$ unless $\Lambda=0$, the symmetry condition is satisfied automatically.

\section{Estimation of $\Rey_s$ for case I}
\label{sec:critical}

As inferred from Fig.~\ref{fig:critical}, to estimate the first critical Reynolds number $\Rey_s$, we can introduce the following regular perturbation series,
\begin{equation}\label{stokesexp}
    f(\xi)=\sum_{n=0}^\infty \Rey^n f_n(\xi) \quad \Rightarrow \quad   f^2(\xi) = \sum_{n=0}^\infty \Rey^n g_n(\xi), \quad g_n(\xi) = \sum_{k=0}^n f_k f_{n-k}.
\end{equation}
At leading order, the solution is found to be
\begin{equation} \label{firstO}
    f_0=-\left(1-\frac{\xi}{\xi_w}\right)^2.
\end{equation}
This solution is simply the negative of that of Schneider's~\cite{schneider1981} when $\Rey =0$, although this symmetry is broken in higher-order terms due to  inertial effects. The first-order solution is found to be
\begin{equation}\label{secondO}
    f_1 = \frac{\xi}{\xi_w^4}\left[2 (2\xi_w-1)(\xi-1)\ln \frac{1-\xi_w}{1-\xi} - (\xi_w-\xi) (3\xi_w +\xi -2)\right], 
\end{equation}
whereas higher order approximations may be obtained from the linear equation
\begin{equation} \label{higherO}
    \xi(1-\xi) f_n' - (1-2 \xi) f_n + g_{n-1} = - \frac{2f_{n-1}'(0)}{\xi_w}\xi (\xi_w-\xi), \quad n\ge 2.
\end{equation}
The second derivatives, necessary for evaluation of critical conditions are given by
\begin{equation}
    f_0''(\xi_w) = -\frac{2}{\xi_w^2}, \quad f_1''(\xi_w) = \frac{2}{\xi_w^2(1-\xi_w)}, \quad \text{and} \quad f_n''(\xi_w) = \frac{2f_{n-1}'(0)}{\xi_w(1-\xi_w)}, \quad n\ge 2,
\end{equation}
that follows directly from~\eqref{fwwfa}.  For practical purposes, we could assume convergence of partial sum for the expansion~\eqref{stokesexp} terminating at a finite value $n=N$. This is permitted since the range of Reynolds number that are required for the current problem is not sufficiently large for things to be otherwise.

Now if we impose the condition $f''(\xi_w)= f_0''(\xi_w) + \Rey_s f_1''(\xi_w) + \Rey_s^2 f_2''(\xi_w)+ \cdots +\Rey_s^N f_N''(\xi_w)=0$ to obtain the critical Reynolds number, we find for $N=1$ and $N=2$, 
\begin{align}
    \Rey_s &= 1-\xi_w,\\
    \Rey_s &= \frac{-1 + \sqrt{1+4\xi_w f^\prime_1(0)(1-\xi_w)}}{2\xi_w f^\prime_1(0)},
\end{align}
respectively. For small $\alpha$, $N$ becomes large. The prediction for three different values of $N=1,2,10$ are shown as dash-dotted curves in Fig.~\ref{fig:critical}.

\section{Boundary-layer structure for case II with $M\leq 1/\xi_w$}
\label{sec:boundarylayer}

 \begin{figure}
\centering
\includegraphics[scale=0.55]{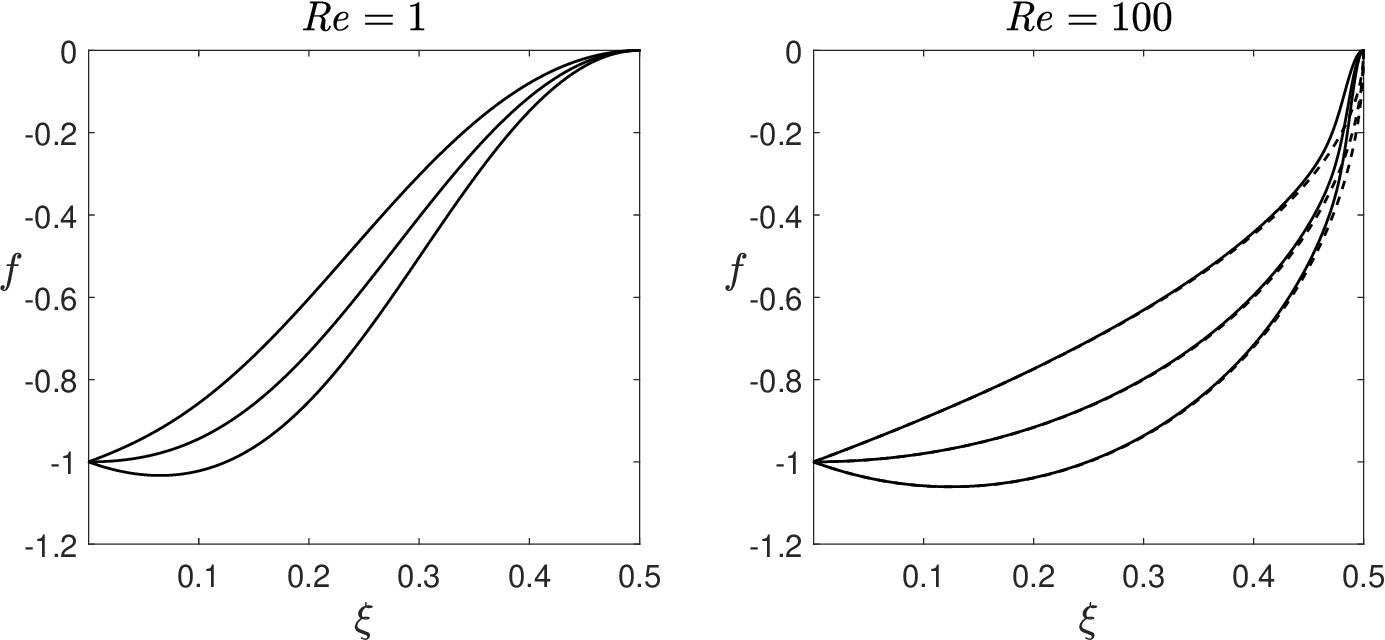}
\caption{Numerical solutions for case II when $\alpha = \pi/2$ and $Re = 1$ (left) and $Re = 100$ (right). Each plot contains curves for $f'(0) = M =-1,0,1$. Dashed lines in the right-hand plot correspond to the potential solution \eqref{potential}.  }
\label{fig:caseIIsol}
\end{figure}

As discussed in \S~\ref{sec:general}, when $M\leq 1/\xi_w$, the flow never separates from the wall and is characterized by pure outflow. Numerical solutions for $\xi_w=1/2$ for two values of $\Rey$ are shown in Fig.~\ref{fig:caseIIsol} for $M=-1,0,1$, all of which are less than $2$. From the right-hand plot, it can be seen that the flow develops an abrupt change in the velocity gradient near the wall indicating the existence of a thin boundary layer near $\xi = \xi_w$ discussed here.

As $\Rey \to \infty$, the potential flow solution can be obtained by solving equation~\eqref{fourthorder} directly after neglecting the terms corresponding to viscous forces, to yield
 \begin{equation}
     f = - \left[1-\frac{\xi^2}{\xi_w^2}- \frac{2M\xi}{\xi_w}(\xi_w-\xi)\right]^{1/2}. \label{potential}
 \end{equation}
 These solutions, which satisfy the no-penetration condition, are plotted in the right-hand side plot of Fig.~\ref{fig:caseIIsol} as dashed curves, agree well with the numerical results except near $\xi=\xi_w$.
 
 However, this potential flow solution induces an infinite slip-velocity at the wall and therefore, the second derivative inside the boundary layer must be very large. This is not a surprising result since we can already see from the formula~\eqref{fwwfa} that $f''(\xi_w)\sim\Rey$ as $\Rey$ becomes large. In practical applications, specifically for the case $M=0$, this means that the bounding wall must withstand the huge stress induced on the wall. With the help of the condition $f''(\xi_w)\sim\Rey$ and using \eqref{potential}, the characteristic boundary layer thickness is found to be $\xi_w-\xi\sim \Rey^{-2/3}$.
 
\begin{figure}
\centering
\includegraphics[scale=0.55]{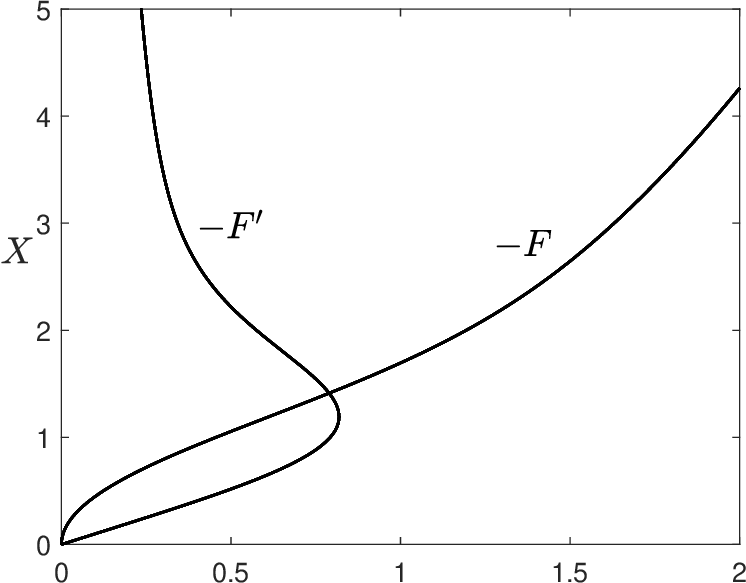}
\caption{Boundary layer profiles for the radial and azimuthal velocities given by \eqref{BLsolution}. }
\label{fig:boundarylayer}
\end{figure}

 The following self-similar scalings
 \begin{equation}
     X = \frac{\Rey^{2/3}(\xi_w-\xi)}{2\xi_w(1-M\xi_w)(1-\xi_w)^{2/3}}, \quad F = \frac{2\Rey^{1/3}(1-M\xi_w) f}{(1-\xi_w)^{1/3}}
 \end{equation}
 yields a parameter-free boundary-layer equation
 \begin{equation}
     F^{iv} - 2 (FF''' + 3 F'F'') =0,
 \end{equation}
 where $F(0)=F'(0)=F(\infty)+\sqrt{X}=0$ needs to be satisfied. Applying the boundary conditions and integrating thrice, we reduce the equation to
 \begin{equation}
     F' - F^2 = - X.
 \end{equation}
 The foregoing equation admits explicit solution in terms of the Airy functions,
 \begin{equation} \label{BLsolution}
     F = -\frac{\sqrt{3} \mathrm{Ai}'(X) + \mathrm{Bi}'(X)}{\sqrt{3} \mathrm{Ai}(X) + \mathrm{Bi}(X)}.
 \end{equation}
 Graphical representations of $F(X)$ and $F'(X)$ representing the radial and azimuthal velocity inside the boundary layer are presented in Fig.~\ref{fig:boundarylayer}. The radial velocity experiences a overshoot inside the boundary layer corresponding to the smallest value of $F'=-0.8186$ that occurs at $X=1.1917$.

\section{Concluding Remarks}

A class of self-similar solutions to the Navier-Stokes equations with distributed half-line-source singularities has been discussed. Two cases, the first in which a symmetry condition was applied and the second in which the radial velocity along the axis was prescribed, were shown to be mutually exclusive. The analysis reveals the complex nature of conically similar flows, including non-uniqueness and non-existence of solutions for a given $\Rey$. The particular case $M=0$ in case II is of central interest from an experimental point of view, since it corresponds to flow injected outwards radially through a thin tube of finite flow rate that necessitates $v_\theta$ to be large and $v_r$ to be zero at the axis. Other applications can also be found in modelling combustion problems; for instance, the diffusion flame surrounding a thin solid fuel rod at the axis in an ambient of oxidizer can be modeled as a distributed line source with outward velocity induced by the effects of thermal expansion. For distances much larger than the radius of the fuel rod the surrounding motion is given by the solution in this paper, which may serve as appropriate boundary condition in studying the details of the fuel burning process. This was the approach taken by~\cite{michaelis2005} to model the velocity displacement induced by the trailing diffusion flame appearing behind a partially-premixed propagating flame front.

\section{Acknowledgments}

The authors would like to express their gratitude to Professors Antonio L S\'anchez and Forman A. Williams.

\bibliographystyle{unsrt}
\bibliography{references}

\end{document}